\documentclass[manuscript]{aastex}

\slugcomment{Submitted to ApJ Letters}

\shorttitle{An Age Spread in NGC 6791}
\shortauthors{Twarog, Carraro, Anthony-Twarog}

\begin{document}


\title{EVIDENCE FOR EXTENDED STAR FORMATION IN THE OLD, METAL-RICH OPEN CLUSTER, NGC 6791?}


\author{Bruce A. Twarog}
\affil{Department of Physics and Astronomy, University of Kansas, Lawrence, KS 66045-7582, USA}
\email{btwarog@ku.edu}

\author{Giovanni Carraro\altaffilmark{1}}
\affil{ESO, Alonso de Cordova 3107, Santiago de Chile, Chile}
\email{gcarraro@eso.org}

\and

\author{Barbara J. Anthony-Twarog}
\affil{Department of Physics and Astronomy, University of Kansas, Lawrence, KS 66045-7582, USA}
\email{bjat@ku.edu}


\altaffiltext{1}{On leave from Dipartimento di Astronomia, Universit\'a di Padova,
Vicolo Osservatorio 3, I-35122, Padova, Italy}


\begin{abstract}
NGC 6791 is an old, metal-rich star cluster normally considered to be a disk open cluster.
Its red giant branch is broad in color yet, to date, there is no evidence for a metallicity spread
among its stars. The turnoff region of the main sequence is also wider than expected from
broad-band photometric errors. Analysis of the color-magnitude diagram reveals a color gradient 
between the core of the cluster and its periphery; we evaluate the potential explanations for this trend. 
While binarity and photometric errors appear unlikely, reddening variations across the face of
the cluster cannot be excluded. We argue that a viable alternative explanation for this color trend 
is an age spread resulting from a protracted formation time for the cluster; the stars 
of the inner region of NGC 6791 appear to be older by $\sim$1 Gyr on average than those of the outer region.
\end{abstract}


\keywords{open clusters: general --- open clusters: individual(NGC 6791)}

\section{Introduction}
Star clusters have long been promoted as ideal tests of stellar evolution because of the
homogeneity in age and chemical composition displayed by their stars; $\omega$ Cen is invariably cited as
the iconic exception that proves this rule. However, the last three decades have witnessed
an increasingly rapid reversal of this rule thanks to improved and expanded spectroscopic and
photometric databases supplying definitive evidence for mixed and sometimes discrete sub-populations in abundance, 
age, or both within the same cluster \citep{PI09}. Since this evidence has been compiled for either Milky Way globular clusters or
the rich, intermediate-age open clusters of the Magellanic Clouds, it suggests that cluster environment -- high total mass 
or stellar density -- is a crucial factor in creating and retaining a mixed population. It would also imply that most 
Milky Way open clusters should display a high degree of chemical and age homogeneity.  

If the environmental scenario is valid, support might exist among the Milky Way open clusters that most resemble 
those of the Magellanic Clouds in mass. Two obvious candidates at intermediate age are NGC 7789 and NGC 2158.
For these clusters, the modest evidence to date provides little support for an internal spread
in age or composition (see, e.g. \citet{MS09}), with the color spread in the color-magnitude diagram (CMD) of NGC 2158 
being a possible byproduct of variable reddening \citep{CA02}. 

A somewhat less obvious candidate is the open cluster NGC 6791. This cluster is unique in the Milky Way, being the most 
massive old open cluster at an age of $\sim$ 8 Gyr, and extremely metal-rich ([Fe/H] $\sim$ +0.40, \citep{CA06, OR06, AT07, BO09}).
Moreover, its extended survival on an orbit which carries it well within the solar circle suggests a system of significantly higher 
mass at the time of its formation \citep{CA06}. Since the CMD-based age spread estimated for populous clusters is usually less than 
the 0.7 Gyr found for NGC 419 by \citet{RU10}, the CMD of NGC 6791 may have resembled that of the Magellanic Cloud clusters 7 Gyrs ago.

While the red giant branch of NGC 6791 is broad in color \citep{JA84, ST03}, as with NGC 7789 and NGC 2158, 
no statistically significant evidence for a metallicity spread within the cluster has been reported so far. 
Although never explicitly mentioned, the upper main sequence (MS) and the turnoff (TO) region of its CMD are broad as well,
an observation in most clusters usually attributed to photometric errors, variable reddening and/or binaries.
We show in this Letter that a highly plausible, though not exclusive, explanation for the MS/TO broadening is an age spread between 
the inner and the outer regions of the cluster, implying possible detection of prolonged, if not discrete, star formation activity 
in NGC 6791.

\section{CMD: Structure and Analysis}
We make use of the $BVI$ photometric database compiled by \citet{ST03}(ST03), complemented by a proper-motion analysis of NGC 6791 
supplied by Dr. Kyle Cudworth (private communication, 2008). Based upon over 1700 CCD frames and covering $\sim$19 
arcmin squared, the photometry of ST03 is of exceptionally high precision to more than 4 mag below the TO. Merger of
the two data sets generated a catalog of 1093 stars with membership probability above 80$\%$ and photometric errors below 0.050 mag in
$B$ and 0.020 mag in $V$ and $I$. The observational data are shown in the CMD's on the left of Fig. 1. For comparison on the right are
synthetic CMD's based upon the isochrones of \citet{GI05, MAR08} for an age of 8.5 Gyr and [Fe/H] = 0.4, with a 
binary percentage of $30\%$. The synthetic CMD has been blurred based upon the quoted photometric errors from ST03. Qualitatively, 
it is apparent that the observed CMD exhibits greater scatter than predicted by the synthetic diagrams. A more quantitative 
illustration is provided by the inset histograms as a function of color for stars between $V$ = 17.6 and 18.0; a small color 
shift has been applied to the histograms with increasing $V$ to account for the mild slope of the TO. The primary 
peak from single stars has approximately twice the fwhm in $B-I$ as the synthetic CMD and is separated in color from the binary 
contribution by an amount which precludes composite stars as the source of the scatter.

Binaries and photometric errors aside, scatter in CMD's can have multiple origins - reddening, age, and metallicity,
to name a few. However, a key element for NGC 6791 emerged as a byproduct of a discussion of its distance based upon
MS-fitting to nearby field stars \citep{TW09}. Fig. 2 illustrates the point. The cluster sample, centered
on ST03 8075, was divided radially into an inner core of radius 150 pixels on the WEBDA scale ($\sim 2\arcmin$) containing
381 stars and an outer region with the remaining 712 stars. The CMD of Fig. 1 is replotted at left with the core stars
in red and the outer region in green. It is evident that the two regions generate slightly different CMD's. To emphasize this, 
the middle panel shows a fit of the core CMD with an isochrone \citep{GI05, MAR08} of age 8.5 Gyr and [Fe/H] = 0.4, 
shifted by $E(B-I)$ = 0.30, and $(m-M)$ = 13.40. In a differential context the exact parameters are not important but 
the isochrone is clearly an excellent match to the data; if anything, the isochrone delineates the bluer edge of the
vertical TO region. 

By contrast, the same isochrone superposed upon the CMD of the outer region (right panel) demonstrates that between $V$ = 18.3 and
the base of the giant branch, the outer region has a bluer TO and brighter subgiant branch, indicative of a younger age,
all other parameters being equal. The color histograms for each region, derived in the same manner as in Fig. 1, confirm that 
the MS broadening is a result of the superposition of two narrower but offset distributions. 

Two key points should be emphasized. First, for stars fainter than the vertical TO, the color
offset disappears and the unevolved MSs are indistinguishable. Second, while we have used the $V, B-I$ diagram to illustrate the
separation to optimal effect, the color separation is apparent with either the $V, B-V$ CMD or the $V, V-I$ CMD, with the 
$B-I$ trend being the sum of the two offsets. Given these constraints, we can now evaluate the likelihood that the offsets are caused 
by photometric errors, reddening, and/or age.

The internal precision of the ST03 photometry is exceptional but this doesn't preclude the possibility of radially-dependent, systematic 
shifts in color. To test this, we compared the $BV$ photometry of ST03 with that of \citet{KR95}, the next most accurate
photometric database covering approximately the same area, derived using a reduction and calibration procedure independent of ST03. $B-V$
was adopted as the color index due the lack of $I$ photometry in the \citet{KR95} survey.

Fig. 3 shows the residuals in $B-V$, in the sense (ST03 - KR95), for all stars brighter than $V$ = 20.0 as a function of radial
position in pixels on the coordinate scale of ST03. Stars with absolute residuals larger than 0.15 mag have been excluded from the analysis. 
The vertical bar illustrates the breakpoint defining inner versus outer region in
Fig. 2. The filled circles show the mean residuals with standard deviations in annuli 100 pixels wide. While there is evidence
that the $B-V$ photometry of ST03 for the inner region is slightly redder than the outer, compared to the system of \citet{KR95}, 
the difference (+0.0070 $\pm$ 0.0015 mag) is too small to produce the color shift as defined by the $V, B-V$ CMD. Even more important,
because the magnitudes are independently calibrated in each filter, a color gradient would only arise if there is a radially-dependent 
offset in each of the individual calibrations which coincidentally combined to produce color offsets which mimicked coherent color/temperature
changes in $B-V$, $V-I$, and $B-I$, but only for stars from the subgiant branch to the TO region. While it cannot be excluded, this
seems somewhat implausible.

Finally, it should be recognized that for area coverage, filter coverage, and especially internal precision, no published photometry for 
NGC 6791 even comes close to \citet{ST03}. Attempts to prove or disprove the form of CMD structure illustrated in Fig. 2 using any other
published source at best will lead to marginal differences, as exemplified by Fig. 3, with no means of determining which dataset
defines the ``correct'' system. For example, a similar comparison was attempted using the $B-I$ data of
\citet{MO94}, producing no evidence for a radial gradient, but the photometric scatter in the latter dataset severely reduced the
statistical significance of the null result.

For an alternative independent check of the reality of the color gradient, the stars defining 
the central peak distributions in a $\pm$0.02 mag range in the histograms of Fig. 2 were identified and compiled using the 
data of ST03 and the intermediate-band photometry of
\citet{AT07}. The mean color difference in $B-I$ for $V$ = 17.6 to 18.0 between the inner and outer regions is 0.0196 $\pm$ 0.0023 (sem). 
(It should be noted that this estimate is a lower bound on the shift required to optimally align the inner and outer region CMD's. 
The CMD for stars above $V$ = 17.6 exhibit the greatest differential
between the inner and outer zones, but the effect is a combined offset in both color and magnitude due to the curvature of the TO
and thus more difficult to quantify than a simple color shift using the approximately vertical zone at the TO.) 
For the intermediate-band color index, $b-y$, the mean difference for the same stars is 0.0137 $\pm$ 0.0038 (sem), consistent with a true
color difference between the two samples but barely a 3.5-sigma effect. The 
intriguing result is that the differentials in $m_1$ and $hk$, the metallicity indicators, both imply that the core region is 
more metal-poor than the outer region by between 0.2 and 0.3 dex, though the statistical significance of this differential is even weaker 
than that of $b-y$ and supplies further evidence of the challenge in trying to identify the origin of the color change. 
Note that the color shift implied by this potential metallicity difference, all things being equal, should make the
core bluer than the outer region.

If photometric errors are not the solution, the next option is variable reddening. The value of this solution is that it helps explain
the apparent lack of a color differential for the unevolved MS. With $\Delta$$E(B-I)$ = 0.030, $\Delta$$E(B-V)$ = 0.013, 
and $\Delta$$A_V$ = 0.040. 
For a change of $B-I$ = +0.03 on the unevolved MS, $V$ changes by +0.09 mag. Thus, the impact of reddening on the scatter in the 
unevolved MS is cut almost in half to 0.017 mag. The full impact is visible at the vertical TO because the distribution
of points is almost orthogonal to the reddening vector. However, the second location in the CMD where the impact should be at least as
noticeable is the vertical giant branch. The inner region giant branch does not appear to separate from the outer region as well as the
TO, if at all, but the scatter in color for both regions is at the same level as the expected shift. The base of the vertical giant branch
does extend redder for the inner core compared to the outer region. However, the morphology difference between the branches leads to
a superposition of both sets as one moves up the giant branch to $V$ = 17.25. Therefore, the evidence for excluding reddening remains
inconclusive. If it is the cause of the offset, the implication is that the stars within the cluster core are reddened more than the
outer region by $\Delta$$E(B-V)$ between 0.010 and 0.015 mag.

\section{Discussion and Conclusions}
We have argued that photometric error can be excluded as the source of the distributions seen in Fig. 2; metallicity effects, if they exist at all, appear to work in the wrong direction.  Having presented arguments questioning the exclusive
role of variable reddening as the culprit, we now consider the viability of an age spread.
The argument is demonstrated in Fig. 4, identical to Fig. 1 except that the synthetic CMD has been computed allowing for the inclusion
of an age spread of 1 Gyr. Clearly an age spread of this size can almost completely account for the MS broadening. This solution has the
added effect of removing any differential for stars on the unevolved MS since the age impact on the CMD will
be increasingly apparent as one moves up the MS to the evolved stars in the TO region, with the maximum separation
at the top of the TO. The color shift in the giant branches caused by this age offset would be less than half the size of the
TO shift and impossible to detect within the scatter of the giant branch.

Should age be the predominant explanation for the broadened turnoff in the CMD of NGC 6791, it implies that the cluster formed during an 
extended star formation episode lasting $\sim$ 1 Gyr, making it similar to 
several Magellanic Cloud clusters, except for the long-standing anomaly of an extremely high metal content for this older cluster. Among 
open clusters, only NGC 6253, with less than half the age of NGC 6791, approaches the metallicity of NGC 6791 \citep{AT10}. 
This result, if correct, enhances the unique position of NGC 6791, as defined by the extreme combination of age, metallicty, and kinematics,
among the Galactic open cluster population; we remind the reader that although NGC 6791 lies within the solar circle in the galactic disk, 
it has an unusually eccentric orbit \citep{CA06}. Collectively, these properties might be an indication of an external origin for NGC 6791.

\acknowledgments
It is a pleasure to acknowledge the significant improvement in this paper resulting from the thoughtful comments of the referee.
Extensive use was made of the WEBDA database maintained by E. Paunzen at the University of Vienna, Austria (http://www.univie.ac.at/webda). 
BAT acknowledges ESO support of a visit to the Vitacura facility where this collaboration was initiated. GC acknowledges ESO DGDF 
support during a visit to the University of Kansas where the paper was prepared.

\clearpage

\begin{figure}
\epsscale{.80}
\plotone{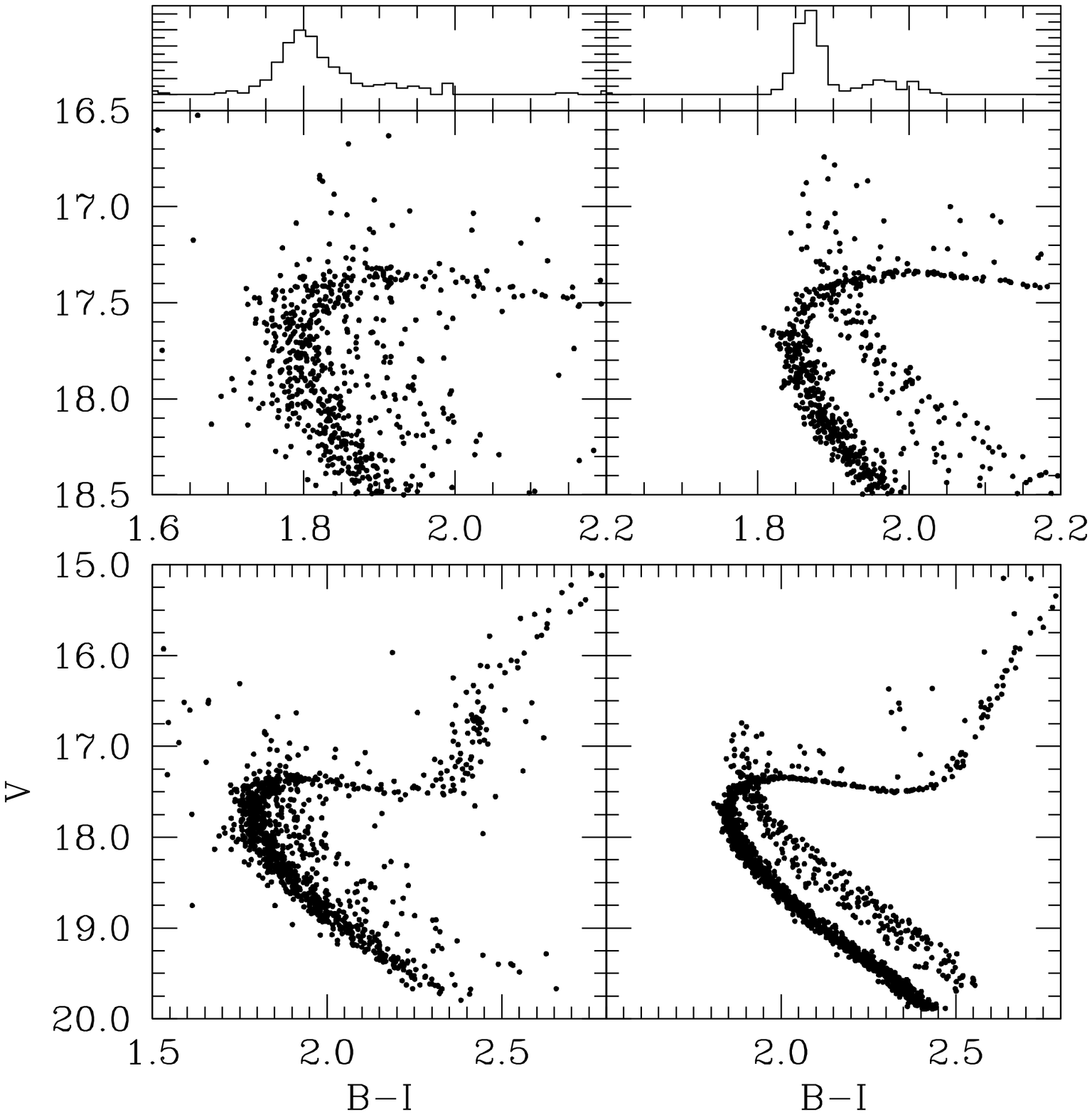}
\caption{Lower left panel: CMD based upon ST03 photometry for stars with proper-motion membership above $80\%$. Lower right: Synthetic 
CMD for an age of 8.5 Gyr, with $30\%$ binaries and observational errors included. Upper panels: Expanded TO view of the two lower panels.}
\end{figure}

\clearpage

\begin{figure}
\epsscale{.80}
\plotone{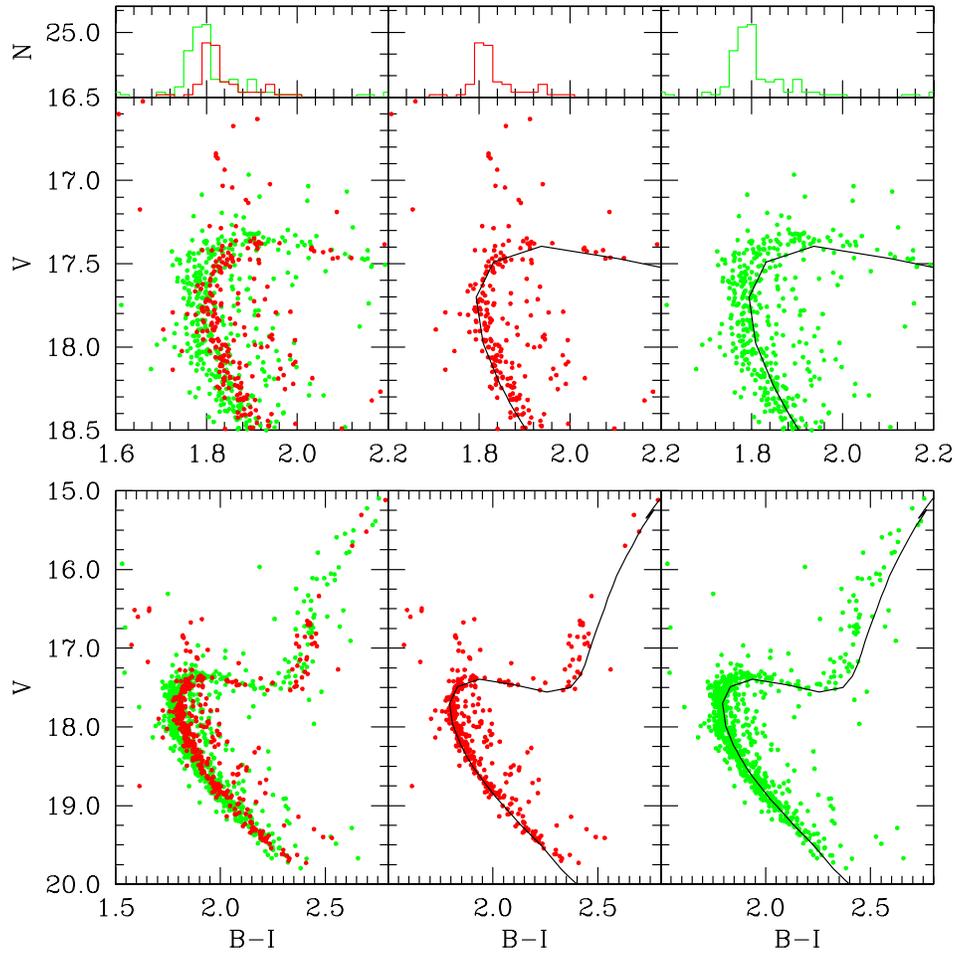}
\caption{ Left panel: Inner (red) and outer (green) region CMD. Middle panel: Isochrone fit (solid line) to the inner cluster CMD.
Right panel: Fit defined by the inner cluster (middle panel) superposed upon the CMD for the outer region.}
\end{figure}

\clearpage

\begin{figure}
\includegraphics[angle=270,scale=0.6]{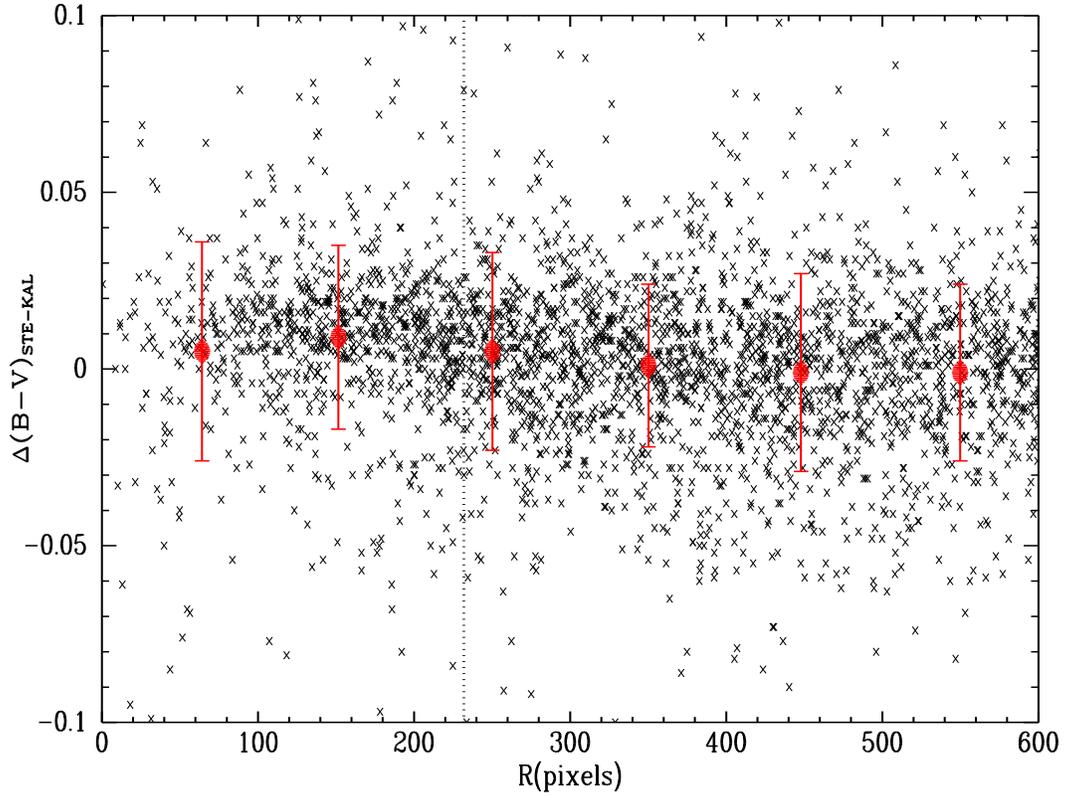}
\caption{ Residuals in $B-V$ between STO3 photometry and \citet{KR95} as a function of radial position. The vertical band defines the
boundary between the inner and outer cluster. Filled points show the mean residuals with standard deviations in annuli 100 pixels wide.}
\end{figure}

\clearpage

\begin{figure}
\epsscale{.80}
\plotone{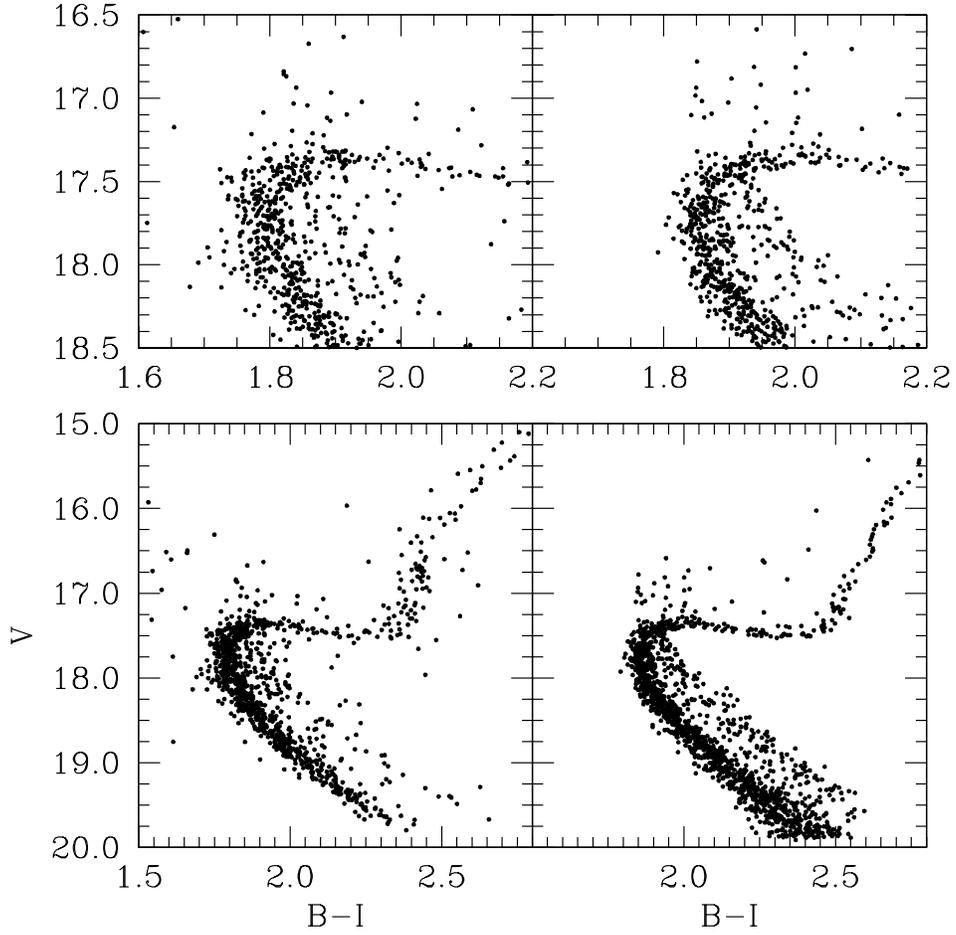}
\caption{ Same as Fig. 1 with the NGC 6791 CMD compared to a synthetic CMD constructed with an extended period of star formation 
lasting  1 Gyr.}
\end{figure}

\end{document}